\documentclass[useAMS,usenatbib,onecolumn]{mn2e}

\usepackage{psfig}
\usepackage{colordvi}
\usepackage{ulem}

\newcommand{\myemail}{agolATastro.washington.edu}
\begin{document}

\title[Transit timing variations]{On detecting terrestrial planets with timing of giant planet transits}

\author[Agol et al.]{Eric Agol$^1$\thanks{\myemail}, Jason Steffen$^1$, 
Re'em Sari$^2$ and Will Clarkson$^3$\\
$^1$Astronomy Department, University of Washington, Box 351580, Seattle, WA 98195 \\
$^2$Theoretical Astrophysics, MS 130-33, Caltech, Pasadena, CA 91125\\
$^3$Department of Physics and Astronomy, The Open University, Milton Keynes MK7 6AA, UK}


\maketitle
\begin{abstract}
The transits of a planet on a Keplerian orbit occur at time
intervals exactly equal to the period of the orbit. If a second planet
is introduced the orbit is not Keplerian and the transits are no
longer exactly periodic.  We compute the magnitude of these variations
in the timing of the transits, $\delta t$. We investigate analytically several
limiting cases: (i) interior perturbing planets with much smaller
periods; (ii) exterior perturbing planets on eccentric orbits with
much larger periods; (iii) both planets on circular orbits with
arbitrary period ratio but not in resonance; and (iv) planets on initially
circular orbits locked in resonance.  Using subscript ``out'' and
``in'' for the exterior and interior planets, $\mu$ for planet to star
mass ratio and the standard notation for orbital elements, our
findings in these cases are as follows: (i) Planet-planet
perturbations are negligible. The main effect is the wobble of the
star due to the inner planet, therefore $\delta t \sim
\mu_{in}(a_{in}/a_{out})P_{out}$. (ii) The exterior planet changes the
period of the interior one by $\mu_{out} (a_{in}/r_{out})^3P_{in}$. As
the distance of the exterior planet changes due to its eccentricity,
the inner planet's period changes. Deviations in its transit timing
accumulates over the period of the outer planet, therefore $\delta t
\sim \mu_{out}e_{out}(a_{out}/a_{in})^3P_{out}$.  (iii) Half way
between resonances the perturbations are small, of order $\mu_{out}
a_{in}^2/(a_{in}-a_{out})^2P_{in}$ for the inner planet (switch
``out'' and ``in'' for the outer planet).  This increases as one gets 
closer to a resonance. (iv) This is perhaps the most interesting case 
since some systems are known to be in resonances and the perturbations 
are the largest. As long as the perturber is more massive than the 
transiting planet, the timing variations would be of order of the period
regardless of the perturber mass! For lighter perturbers, we show that
the timing variations are smaller than the period by the perturber to
transiting planet mass ratio.  An earth mass planet in 2:1 resonance
with a 3-day period transiting planet (e.g. HD 209458b) would cause
timing variations of order 3 minutes, which would be accumulated over
a year. These are easily detectable with current ground-based 
measurements.

For the case of both planets on eccentric orbits, we compute
numerically the transit timing variations for several cases of known
multiplanet systems assuming they were edge-on.  
Transit timing measurements may be used to constrain the
masses and radii of the planetary system and, when combined with
radial velocity measurements, to break the degeneracy
between mass and radius of the host star.
\end{abstract}

\section{Introduction}

The recent discovery of planets orbiting other stars (``exoplanets'')
has opened a new field of astronomy with the potential to address
fundamental questions about our own solar system which we can
now compare with other planetary systems.  The primary mode 
for discovery of
exoplanets has been the measurement of the stellar radial velocities
via the Doppler effect. Currently the small reflex motion of the star
due to the orbiting planet can only be detected for $m_{planet}\ga 
10 m_\oplus$ \citep{but04,mca04,san04}.  More recently, a large number of
planetary transit searches are being carried out which
are starting to yield an handful of giant planets 
\citep{cha00,kon03,pon04,kon04,bou04, alo04}, and many more planned 
searches should reap a large harvest of transiting planets in the
near future \citep{hor03}.  Despite these successes, the discovery of 
``terrestrial'' exoplanets, similar in size to the earth, awaits the 
development of several other techniques such as astrometry, space-based
transit searches, microlensing, or direct imaging \citep{per00,cha03,for03,bor03a,
gou04}.

The first transiting planetary system, HD 209458b, was discovered with 
Doppler motions of the primary star \citep{cha00}.  Since the mass of the
planet is degenerate with orbital inclination, the planetary status of
the companion was confirmed since the transits imply it is edge-on.  {\it
HST} observations yielded precision measurements of the transit
lightcurve \citep{bro01}, which made this the surest planetary
candidate around a main sequence star (other than our own).  The ratio
of the planetary radius to the stellar radius can be measured with
extreme precision \citep{man02}.  However, the absolute radii are
uncertain due to a degeneracy between radius and mass of star
\citep{sea03a}: an increase in the mass and radius of the star can
yield an identical lightcurve and period.

This mass-radius degeneracy may be broken 
if there is an additional planet in the system.
About 10 per cent of the stars with known planetary companions have more than 
one planet, while possibly as much as 50 per cent of 
them show a trend in radial velocity indicative of
additional planets \citep{fis01}.
If one or both of the planets is transiting, dynamical interactions
between the planets will alter the timing of the transits
{\citep{dob96,cha00,mir02}.}  A measurement of these timing
variations, combined with radial velocity data, can break the
mass-radius degeneracy. 

Given the dual motivations of detecting terrestrial planets and breaking
the mass-radius degeneracy, we derive analytic and numerical results 
for transit timing variations due to the presence of a second planet.  
We begin our discussion by introducing the three-body system in \S 2.  
The signal from non-interacting planets is calculated first (\S 3)
and then we compute the effects of an eccentric exterior perturbing 
planet with a large period in \S 4.
A derivation of the general transit timing differences for two planets
with circular, co-planar orbits is presented in \S 5.  The case of exact 
mean-motion resonance is analyzed in \S 6.  The case of two eccentric planets 
is considered in \S 7, along with numerical simulations of several known
multi-planet systems (these are not transiting).  We show 
how measurements of the dispersion of transit timings can be used to detect 
a secondary planet in the system (\S 8.1), we compare to other planet-search 
techniques (\S 8.2), and we show how to determine the absolute size and mass of the 
objects in the system (\S 8.3).  Finally, we discuss other effects
we have ignored that an observer should be conscious of (\S 9).

\begin{figure}
\centerline{\psfig{file=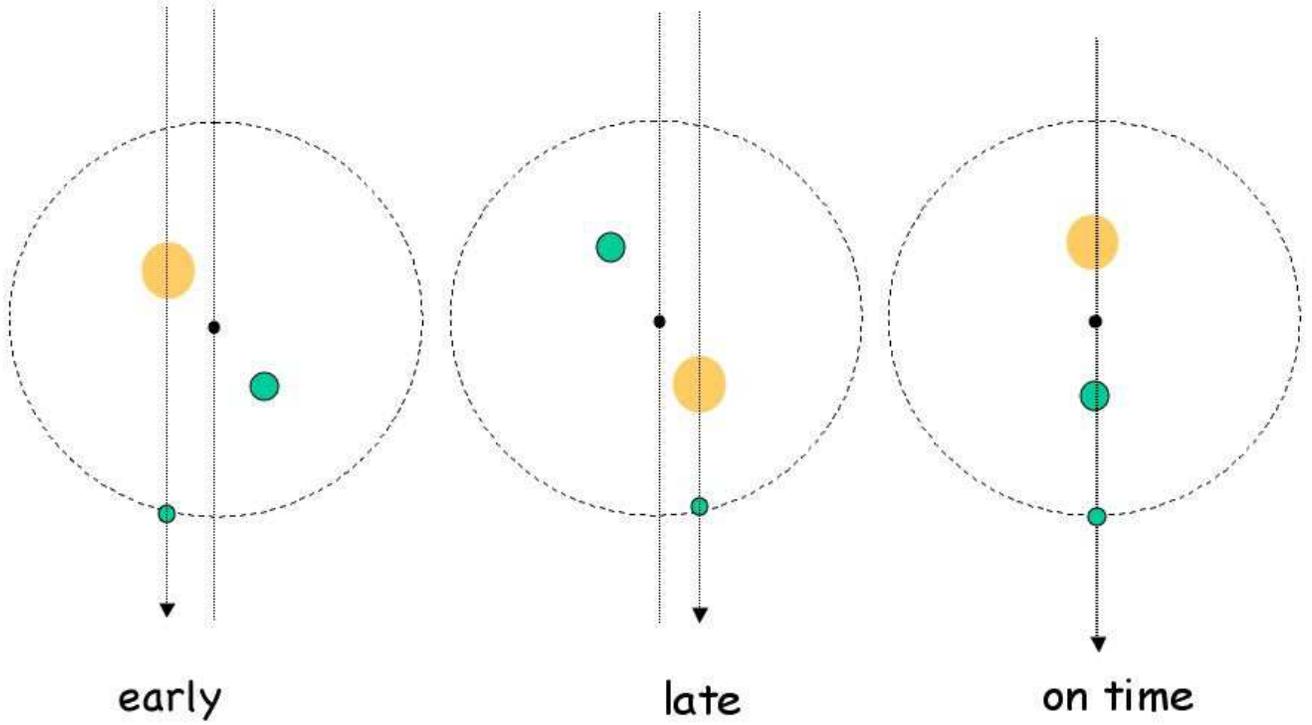,width=\hsize}} 
\caption{Cartoon showing changes in the timing of transit due to
a perturbing planet interior to the orbit of a transiting planet.}
\label{fig1}
\end{figure}

Throughout the rest of the paper we characterize the strength
of transit timing variations as follows.
For a series of transit times, $t_j$, we fit the times assuming
a constant period, $P$.  We compute the standard deviation, 
$\sigma$, of the difference between the nominal and actual times.  
Mathematically,
\begin{equation}
\sigma={1\over N}\left[\sum_{j=0}^{N-1} (t_j-t_0+Pj)^2\right]^{1/2},
\end{equation}
where $P$ and $t_0$ are chosen to minimize $\sigma$.
If the variations are strictly
periodic, then the amplitude of the timing deviation is
simply $\sqrt{2}$ times larger than $\sigma$. 

During the preparation of this
paper a proceedings contribution has appeared by Jean Schneider which
considers several of the effects discussed here \citep{sch03}; however,
we find that Schneider's results are incorrect as he does not consider
the differential force between the star and the transiting planet.  In 
addition, calculations similar to those presented here are being carried 
out by Matt Holman and Norm Murray.

\section{Equations of Motion}

We are studying the 3-body system in which the three bodies have
labels $0,1,2$ and positions ${\bf R}_i, i=0,1,2$ (with an arbitrary
origin).  The exact Newtonian equations of motion are given by
\begin{equation}
\ddot{\bf R}_i = \sum_{j\ne i} G m_j{{\bf R}_j-{\bf R}_i
\over \vert{\bf R}_j-{\bf R}_i\vert^3}.
\end{equation}

Multiplying the equations for each particle by its mass and adding
together, one finds:
\begin{equation}
m_0 \ddot{\bf R}_0+m_1 \ddot{\bf R}_1+m_2 \ddot{\bf R}_2 = 0.
\end{equation}
This is simply a statement that the centre of mass of the system has
no external forces. Since light travel time and
parallax effects are negligible (\S\ref{lighttravel}), the transit problem is
unaffected by the total velocity or position of the centre of mass, so
we set
\begin{equation}
{\bf R}_{cm}\equiv {\sum_{i=0}^2 m_i {\bf R}_i \over \sum_{i=0}^2 m_i} = 0.
\end{equation}
This reduces the differential equations
of motion to two, which we take to be that of the two planets, $R_1$
and $R_2$ (for the two planetary masses). 
We use this system of
equations for numerically solving the equations of motion.  However,
for analytic consideration it is more convenient to write the problem in
Jacobian coordinates
which we discuss next.

The Jacobian coordinate system is commonly used in perturbation theory
for many bodies \citep[see, e.g.][]{mur99, mal93a, mal93b}.  For the
three-body problem, the Jacobian coordinates amount to three new
coordinates which describe (a) the centre of mass of the system; (b)
the relative position of inner planet and the star (the ``inner
binary"); (c) the relative position of the outer planet and the
barycentre of the inner binary (the ``outer binary'').  To distinguish
from the body coordinates, we denote the Jacobian coordinates with a
lower case ${\bf r}_i$.  The Jacobian coordinates are
\begin{eqnarray}
{\bf r}_0 &=& {\bf R}_{cm} = 0,\cr
{\bf r}_1 &=& {\bf R}_1 - {\bf R}_0,\cr
{\bf r}_2 &=& {\bf R}_2 - {1 \over m_0+m_1}\left[m_0{\bf R}_0+m_1{\bf R}_1\right].
\end{eqnarray}
Using  $\mu_i = m_i/M \sim m_i/m_0$, where $M=\sum_{i=0}^2 m_i$,
the equations of motion may be rewritten in Jacobian coordinates,
\begin{eqnarray}\label{jaceq}
\ddot{{\bf r}}_1 &=& -{Gm_0\over 1-\mu_1}{{\bf r}_1 \over 
 r_1^3} -GM\mu_2 {{\bf r}_1 - {\bf r}_{21}
\over \vert {\bf r}_1 - {\bf r}_{21}\vert^3}
- GM\mu_2 {{\bf r}_{21} \over
r_{21}^3 },\cr
\ddot{{\bf r}}_2 &=& -{Gm_0\over 1-\mu_2}{{\bf r}_{21} 
\over r_{21}^3}  - GM\mu_1 {{\bf r}_{21} - {\bf r}_1 \over
\vert{\bf r}_{21} - {\bf r}_1\vert^3},
\end{eqnarray}
where 
${\bf r}_{21}=\mu_1{\bf r}_1+{\bf r}_2={\bf R}_2-{\bf R}_0$.

\section{Non-interacting planets: Perturbations due to interior planet on a small orbit }

Throughout the rest of the paper we make the approximations that
(a) the orbits of both planets are aligned in the same plane; 
(b) the system is exactly edge-on, that is, the inclination angle
is 90$^\circ$.
We also approximate the planet
and star as spherical so that the transit is symmetric with a
well-defined midpoint.

If we take the limit as $\mu_1,\mu_2 \rightarrow 0$ in equation 
(\ref{jaceq}), then the
orbits of the planets follow Keplerian trajectories
with the equations of motion
\begin{eqnarray}\label{jackepler}
\ddot{{\bf r}}_1 &=& -Gm_0{{\bf r}_1 \over
 r_1^3},\cr
\ddot{{\bf r}}_2 &=& -Gm_0{{\bf r}_2 \over
 r_2^3}.
\end{eqnarray}
This approximation requires that the periapse of the
outer planet be much larger than the apoapse of the inner planet,
$(1-e_2)a_2\gg (1+e_1)a_1$ where $a_1, a_2$ are the semi-major axes of
the inner and outer binary and $e_1, e_2$ are the eccentricities.
In this case, the inner binary orbits about its barycentre which in
turn orbits about the barycentre of the outer binary but there is no
perturbation to the relative motion of the inner binary due to
gravitational interactions.  Timing variations that arise are
simply due to the reflex motion of the star (as shown in Figure \ref{fig1}).

The simplest case to consider is that in which both the inner and
outer binary are on approximately circular orbits.  The transit occurs
when the outer planet is nearly aligned with the barycentre of the
inner binary and its motion during the transit is
essentially transverse to the line of sight.
The inner planet displaces the
star
from the barycentre of the inner binary by an amount
\begin{equation}
x_0=-a_1\mu_1\sin{\left[2\pi(t-t_0)/P_1\right]},
\end{equation}
where the inner binary undergoes a transit at time $t_0$ and $P_1$
is the orbital period of the inner binary.  Thus, the timing deviation of
the $m$th transit of the outer planet is
\begin{equation}
\delta t_2 \approx -{x_0 \over v_2-v_0} \approx
-{P_2a_1\mu_1\sin{\left[2\pi(mP_2-t_0)/P_1\right]}
\over 2\pi a_2},
\end{equation}
where $v_i$ is the
velocity of the $i$th body with respect to the line of sight.
Typically $v_0 \ll v_2$, so we have neglected $v_0$ in the second
expression in the previous equation.

Computing the standard deviation of timing variations over many orbits gives
\begin{equation}\label{sigcirc2}
\sigma_2 = \langle (\delta t_2)^2\rangle^{1/2} = {P_2a_1\mu_1
\over 2^{3/2}\pi a_2}.
\end{equation}
Note that if the periods have a ratio $P_2$:$P_1$ of the form 
$q$:$1$ for some integer $q$, 
then the perturbations disappear
since the argument of the sine function is the same each orbit.
Another observable is the duration of the transit, which scales as
\begin{equation}
t_{2} \approx {2R_* \over (v_2-v_0)}.
\end{equation}
This leads to significant variations only if
$v_0 \simeq v_2$, or $\mu_1^2 > a_1/a_2$, which requires a very 
large axis ratio.

More interesting variations occur if either or both planets are
on eccentric orbits.  Because both planets are following approximately
Keplerian orbits, the transit timing variations and duration variations
can be computed by solving the Kepler problem for each
Jacobian coordinate.  Since we are assuming that the planets are coplanar
and edge-on, 4 coordinates each suffice to determine the planetary
positions:  $e_{1,2}$, $a_{1,2}$, $\varpi_{1,2}$ (longitude of pericentre), 
and $f_{1,2}$ (true anomaly).  As in the circular case, the change in the transit timing 
is approximately $\delta t_2 \approx {x_0 / v_2}$.
The position of the star with respect to the barycentre
of the inner binary is
\begin{equation}
x_0 = -\mu_1 r_1 \sin{\left[f_1+\varpi_1\right]}.
\end{equation}
If $a_1\ll a_2$, the outer planet 
is in nearly the same position at the time of each transit 
and its velocity perpendicular to the line of sight is 
\begin{equation}
v_2={2\pi a_2\left(1+e_2\cos{\varpi_2}\right)\over P_2 \sqrt{1-e_2^2}},
\end{equation}
where we have used the fact that $f_2=-\varpi_2$ at the timing of
the transit.  Thus, to first order in $a_1/a_2$
\begin{equation}
\delta t_2 = -{P_2\mu_1 r_1 \sin{[f_1+\varpi_1]}\sqrt{1-e_2^2}\over
2\pi a_2(1+e_2\cos{\varpi_2})}.
\end{equation}

The standard deviation of $\delta t_2$ can be computed analytically
as well.  Over many transits by the outer planet, the inner binary's position
populates all of its orbit provided the planets do not have a period 
ratio that is the ratio of two integers.  Consequently, we find 
the mean transit 
deviation by averaging over the probability that the inner binary
is at any position in its orbit, $p(f_1)=n_1/\dot f_1$ (where $n_1=2\pi/P_1$),
times the transit deviation at that point.  This gives
\begin{eqnarray}
\langle \delta t_2 \rangle &=& {1 \over 2\pi} \int_0^{2\pi} df_1 \delta t_2
p(f_1) \cr
&=& -{3 \over 2} \mu_1 {a_1\over v_2} e_1 \sin{\varpi_1}.
\end{eqnarray}
Since the star spends more time near apoapse, the mean timing grows as $e_1$.
The symmetry of the orbit about $\varpi=0$ and $\pi$ explains the dependence
on $\sin{\varpi_1}$.
A similar calculation gives $\langle \delta t_2^2 \rangle $ and the resulting standard deviation is
\begin{eqnarray}\label{standarddeviation2}
\sigma_2 &=&\left(\langle\delta t_2^2\rangle-\langle\delta t_2\rangle^2\right)^{1/2}\cr
&=&{P_2a_1\mu_1 \sqrt{1-e_2^2}\over 2^{3/2}\pi a_2(1+e_2\cos{\varpi_2})}\left[1-{e_1^2\over 2}
\left(1+\cos^2{\varpi_1}\right)\right]^{1/2}.
\end{eqnarray}
This agrees with equation (\ref{sigcirc2})
in the limit $e_1 \rightarrow 0$.  Averaging again over $\varpi_1$ and
$\varpi_2$, gives
\begin{equation}
\langle\sigma_2\rangle_{\varpi_1,\varpi_2} =
{P_2a_1\mu_1 \left[1-{3\over 4}e_1^2\right]^{1/2}.\over 2^{3/2}\pi a_2(1-e_2^2)^{1/4}}
\end{equation}
Note that an eccentric inner orbit 
reduces $\sigma_2$ because the inner binary
spends more time near apoapse as the eccentricity increases 
thus reducing the variation in position when 
averaged over time.  
As $e_1$ approaches unity for an orbit viewed along the
major axis, $\sigma_2$ reduces to zero since the minor axis approaches
zero leaving no variation in the $x_0$ position.

\section{Perturbations due to exterior planet on a large eccentric orbit}

In this section we include planet-planet interactions and compute 
the timing variations due to the presence of a perturbing planet on an 
eccentric orbit with a semi-major axis much larger than
that of a transiting planet
on a nearly circular orbit.  In this limit, resonances are
not important and the ratio of the semi-major axes can be used
as a small parameter for a perturbation expansion.  A general
formula for this case has been derived by \citet{bor03}.
Here we present a
shorter derivation which clarifies the primary physical effects
for coplanar planets viewed edge-on.

The equations describing the inner binary can be divided into 
a Keplerian equation (\ref{jackepler}) and a perturbing force
proportional to $m_2$.  The perturbing acceleration $\delta \ddot{\bf r_1}$ 
on the inner binary due to the outer planet is given by
\begin{equation}
\delta \ddot{\bf r_1} = -GM\mu_2 {{\bf r}_1 - {\bf r}_{21}
\over \vert {\bf r}_1 - {\bf r}_{21}\vert^3}
- GM\mu_2 {{\bf r}_{21} \over r_{21}^3 }.
\end{equation}
We expand this in a Legendre series
and keep terms up to first order in the ratio of the radii
of the inner and outer orbit,
\begin{equation}
\delta\ddot{\bf r_1} = -{G m_2 \over r_2^3}\left[{\bf r}_1-3{{\bf r}_1\cdot
{\bf r}_2\over r_2^2}{\bf r}_2\right] + {\cal O}(r_1/r_2)^2.
\end{equation}
To compute the perturbed orbital period we must find
the change in the force on the inner binary due to the outer planet
averaged over the orbital period of the inner binary.
Since the inner binary is nearly circular, the angle of the
inner binary is given by  $\theta_1 = f_1 + \varpi_1 \simeq n_1(t-\tau_1)
\varpi_1$, where we have approximated $e_1\simeq 0$.  
Differentiating this with respect to time gives
\begin{equation}
\dot \theta_1 = \dot n_1 (t-\tau_1) + n_1 - n_1\dot \tau_1.
\end{equation}
Now, we write $\dot n_1 = -3n_1/(2a_1)\dot a_1$, and
express $\dot a_1$, $\dot \varpi_1$,
and $\dot \tau_1$ in terms of the radial, tangential, and normal
components of the force \citep[see section 2.9 of][]{mur99}.  Plugging
these expressions into $\dot\theta_1$ gives a cancellation of most 
terms to lowest order in $e_1$, and after setting the normal force to 
zero leaves the remaining term
\begin{equation}
\dot \theta_1 = n_1 \left(1 - {2 a_1^2 \over G(m_0+m_1)}\bar R\right),
\end{equation}
where $\bar R$ is the radial disturbing force per unit mass, $ \bar R = (\delta
\ddot {\bf r}_1 \cdot \hat {\bf r}_1) = {1 \over 2} Gm_2a_1/r_2^3$.

Thus, the presence of the second planet causes a change in the effective 
mass of the inner binary by an amount $-{1 \over 2}m_2(a_1/r_2)^3$, 
which results in a slight increase in the period of the orbit.  
The increase in period would be constant if
the second planet were on a circular orbit.  However, for an eccentric orbit, 
the time variation of $r_2$ induces 
a periodic change in the orbital frequency of the inner binary
with period equal to $P_2$.

Now, the time of the $(N+1)$th transit occurs at 
\begin{eqnarray}\label{eclipsetime}
t_{ec}-t_0&=&\int_{f_0}^{f_0+2\pi N} df_1 \dot \theta_1^{-1} \cr
&=& \int_{f_0}^{f_0+2\pi N} df_1
n_1^{-1} \left[1+{m_2 \over m_0+m_1} \left({a_1\over r_2}\right)^3\right].
\end{eqnarray}
where $f_0$ is the true anomaly of the inner binary at the time of the first 
transit.  Following \citet{bor03}, we change the variable of
integration from $f_1$ to $f_2$, the true anomaly of the outer planet,
\begin{equation}
df_1 = {P_2\over P_1}{r_2^2\over a_2^2(1-e_2^2)^{1/2}}df_2.
\end{equation}
Since we are assuming that the orbit of the outer planet is eccentric, 
$r_2=a_2(1-e_2^2)/(1+e_2\cos{f_2})$, which 
gives the transit time
\begin{equation}\label{eccentricouter}
t_{ec}-t_0=NP_1+{m_2\left(1-e_2^2\right)^{-3/2}\over 2\pi(m_0+m_1)}{P_1^2\over P_2}
 \left(f_2+e_2\sin{f_2}\right),
\end{equation}
where $f_2$ is the true anomaly of the outer binary at the
timing of the $(N-1)$ transit.  The unperturbed $f_2$ includes 
the mean motion, $n_2(t-\tau_2)$, which grows linearly with time.
To find the deviation of the
time of transits from a uniform period, we subtract off this mean
motion as well as $NP_1$ which results in 
\begin{eqnarray}
\delta t_1 &=& \beta
\left(1-e_2^2\right)^{-3/2} \left[f_2-n_2(t-\tau_2)+e_2\sin{f_2}\right]\cr
\beta&=& {m_2 \over 2\pi(m_0+m_1)}{P_1^2\over P_2}.
\end{eqnarray}
This agrees with the expression of \citet{bor03} in the limit $I=1$
(i.e. coplanar orbits).  Remarkably, the timing variations scale
as $a_2^{-3/2}$, which is a much shallower scaling than estimated
by \citet{mir02}, $a_2^{-3}$.  

\begin{figure}
\centerline{\psfig{file=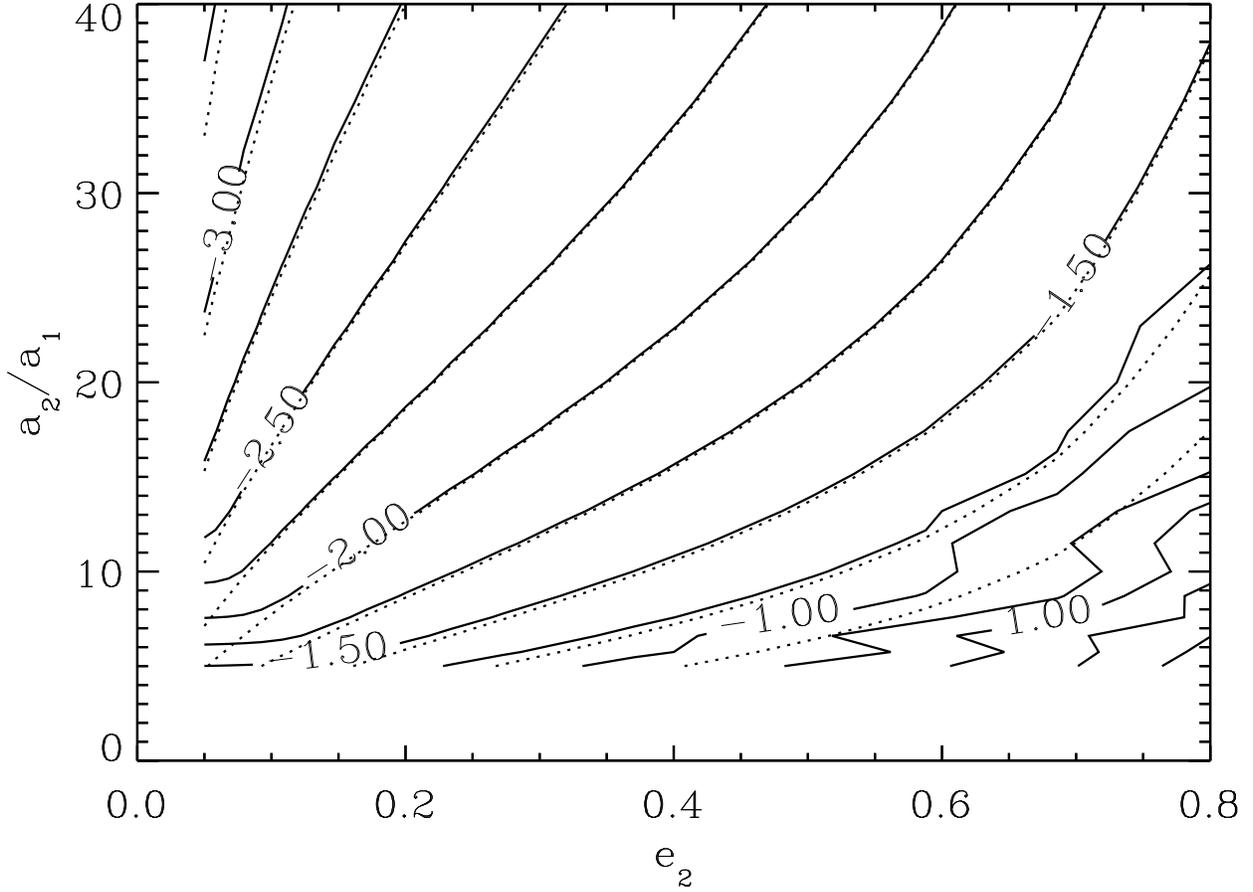,width=\hsize}} 
\caption{Contour plot of the logarithm of the dispersion of the normalized 
timing variations, $\log{(\sigma_1 n_1 \mu_2^{-1})}$, for an inner circular 
planet and an outer eccentric planet (for example, at the -2 contour an
orbit lasting $2\pi$ years with a perturbing planet of mass $10^{-3}
{\rm M_\odot}$
would have a transit timing standard deviation of $10^{-5}$ years, or 
5 minutes). The dotted line is the approximation given in equation 
(\ref{sig2approx}).}
\label{fig2}
\end{figure}

Numerical calculation of the 3-body
problem show that this approximation works extremely well in the 
limit $r_1\ll r_2$ (see Figure \ref{fig2}).  
If $P_1\ll P_2$ and the period ratio 
is non-rational, then over a long time the transits of the inner planet 
sample the entire phase of the outer planet.  Thus, we can compute the
standard deviation of the transit timing variations as in equation 
(\ref{standarddeviation2})
\begin{equation}
\sigma_1 = \langle \delta t_1^2 \rangle^{1/2} = {1 \over 2\pi} \int_0^{2\pi} df_2 \delta t_1^2 p(f_2),
\end{equation}
since $\langle \delta t_1\rangle=0$, where $p(f_2)=n_2/\dot f_2$.
This integral turns out to be intractable analytically, but
an expansion in $e_2$ yields
\begin{equation}\label{sig2approx}
\sigma_1 =  {3\beta e_2 \over \sqrt{2}\left(1-e_2^2\right)^{3\over 2}}
\left[1-{3\over 16}e_2^2-{47\over 1296}e_2^4-{413\over 27648}e_2^6\right]^{1/2},
\end{equation}
which is accurate to better than 2 per cent for all $e_2$.
Figure \ref{fig2} shows a comparison of this approximation with the
exact numerical results averaged over $\varpi_2$ (since there is a
slight dependence on the value of $\varpi_2$).  This approximation
breaks down for $a_2(1-e_2) \la 5 a_1$ since resonances and higher
order terms contribute strongly when the planets have a close approach.
It also breaks down for $e_2 \la 0.05$ since the perturbations to
the semi-major axes caused by interactions of the planets contribute
more strongly than the tidal terms which become weaker with smaller
eccentricity.

\section{Perturbations for two non resonant planets on initially circular orbits}
\label{circularpert}

In this section we estimate the amplitude of timing variations for
two planets on nearly circular orbits.  The resonant forcing terms 
are most important in determining the amplitude, even for non-resonant 
planets.  The planets interact most strongly at conjunction, so the
perturbing planet causes a radial kick to the transiting planet 
giving it eccentricity.  Since the
planets are not exactly on resonance, the longitude of conjunction
will drift with time, causing the kicks to cancel after the longitude
drifts by $\simeq \pi$ in the inertial frame.  Thus, the total amplitude
of the eccentricity grows over a time equal to half of the period of 
circulation of the longitude of conjunction. The closer the planets are 
to a resonance, the longer the period of circulation and thus the 
larger the eccentricity grows.  The change in eccentricity
causes a change in the semi-major axis and 
mean motion.

For two planets that are on circular orbits near a $j$:$j+1$ resonance,
conjunctions occur every $P_{conj}=2\pi/(n_1-n_2)\simeq jP$ (we take
the limit of large $j$ and we ignore factors of order unity).  We define 
the fractional distance from resonance, $\epsilon
= \vert1-(1+j^{-1})P_1/P_2\vert < 1$, 
where $\epsilon=0$ indicates exact resonance.  Then, because the 
planets are not exactly on resonance, the longitude of conjunction 
changes with successive conjunctions and the longitude of conjunction 
returns to its initial value over a period 
$P_{cyc}=Pj^{-1}\epsilon^{-1}$.  The number of conjunctions per cycle 
is $N_c=P_{cyc}/P_{conj}\simeq j^{-2}\epsilon^{-1}$.
Each conjunction changes the eccentricity of the planets by
$\Delta e \sim \mu_{pert} (a/\Delta a)^2$ (using the perturbation equations
for eccentricity and the impulse approximation, where $\mu_{pert}$
is the planet-star mass ratio of the perturbing planet).  Over half a cycle the
eccentricities grow to about $N_c \Delta e \sim \mu_{pert} (1-P_1/P_2)^{-1}
(j\epsilon)^{-1} \simeq \mu_{pert} \epsilon^{-1}$.  

To find the change in the transit timing, we use the orbital
elements to compute the variation in the instantaneous orbital
frequency, $\dot\theta$.  To first order in $e$
\begin{equation}
\dot\theta = {n \left(1+e \cos{f}\right)^2 \over 
\left(1-e^2\right)^{3/2}} 
\approx n_{0} + \delta n + 2 e n_{0}\cos{\left[\lambda-\varpi\right]},
\end{equation}
where $n_{0}$ is the unperturbed mean motion.  There are two terms
which contribute to timing variations:  fluctuations in the mean-motion
and fluctuations due to a non-zero eccentricity.
In the first case, $\delta n$ may be found
by applying the Tisserand relation to the lighter planet (we now use
subscripts ``light" and ``heavy"), resulting in 
$\delta n_{light}/n_{light} \simeq j (\Delta e_{light})^2
\simeq j\mu^2 \epsilon^{-2}$ (where $\mu$ is $\mu_{heavy}$).  
These changes to the period accumulate over an entire cycle, giving
\begin{equation}\label{lightplanet}
\delta t_{light} \simeq \mu^2 \epsilon^{-3} P.
\end{equation}
By conservation of
energy, the fractional change in semi-major axis (or period) of the 
heavy planet is reduced by
a factor of $m_{light}/m_{heavy}$, so that 
\begin{equation}\label{heavyplanet}
\delta t_{heavy}\simeq (m_{light}/m_{heavy}) \mu^2 \epsilon^{-3} P.
\end{equation}

The eccentricity dominated term gives a timing variation of 
\begin{equation}\label{wings}
\delta t \simeq \mu_{pert} \epsilon^{-1} P.
\end{equation}
So, for $\epsilon > \mu^{1/2}$ 
the perturbed eccentricity dominates, but closer to resonance for 
$j^{1/3}\mu^{2/3} < \epsilon < \mu^{1/2}$ the perturbed mean motion dominates 
(this range is the same for both the light and heavy planets, except for
factors of order unity).  For
smaller values of $\epsilon$, the planets are trapped in mean-motion resonance,
which is discussed in the next section.   Half way between
resonances, $\epsilon\simeq j^{-2}$,  so the timing deviation become
\begin{equation}\label{halfres}
\delta t \sim 0.7 \mu_{pert} (a/(a_2-a_1))^2P.
\end{equation}
A more precise derivation in the eccentricity-dominated case
using perturbation theory is given in Appendix \ref{append1}.

So far we have discussed the timing variations for planets nearby
a first order resonance.  For larger period ratios, the eccentricity
of the inner planet grows to $e_{in} \simeq \mu_{out}(P_{in}/P_{out})^2$,
so $\delta t_{in} \sim \mu_{out} (P_{in}/P_{out})^2 P_{in}$.
For an outer transiting planet the motion of the star dominates
over the perturbation due to the inner planet for 
$P_{out}>(2\pi)^{3/4}P_{in}$.

\begin{figure}
\centerline{\psfig{file=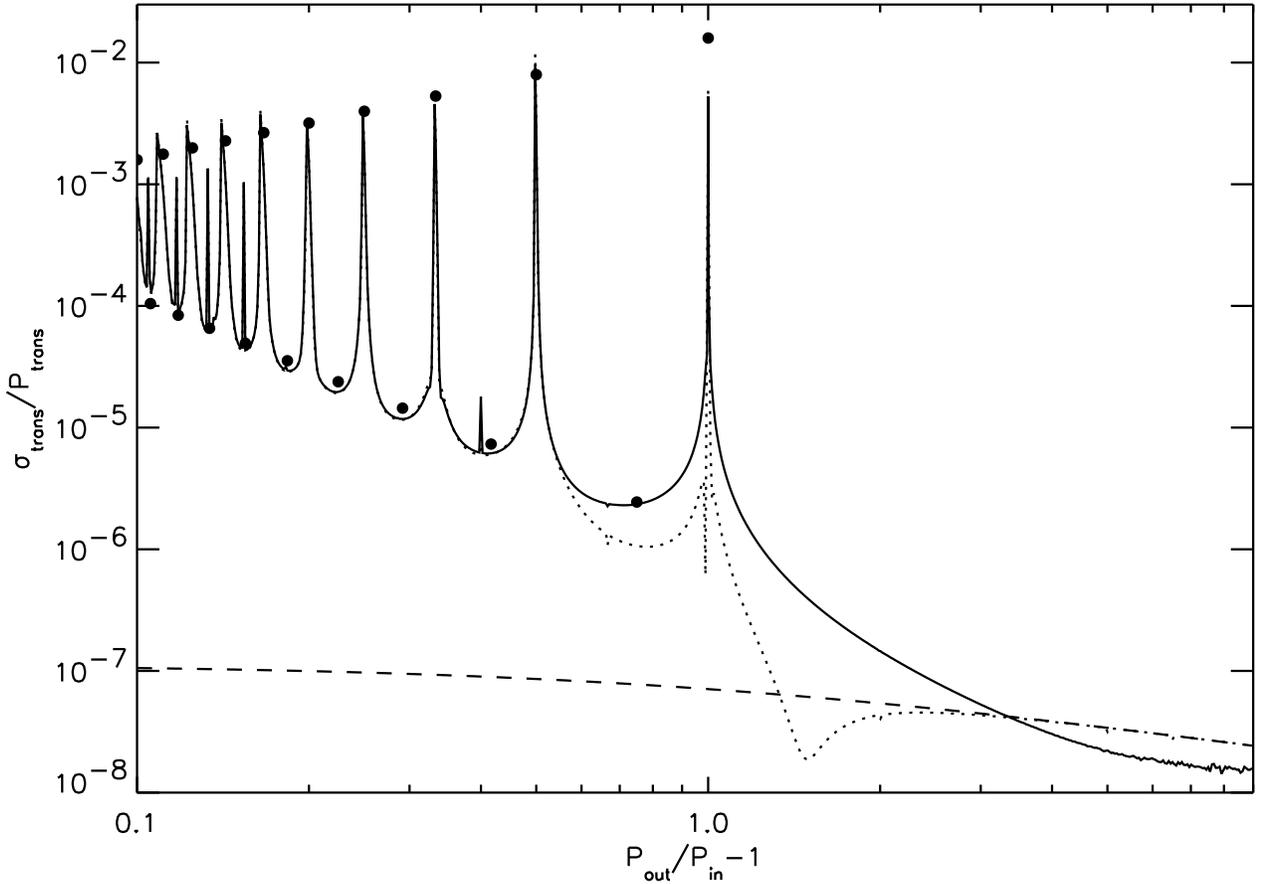,width=\hsize}} 
\caption{Transit timing standard deviation for two planets of mass
$m_{trans}=10^{-5}m_0$ and $m_{pert}=10^{-6}m_0$ on initially
circular orbits in units of the period of the transiting planet.  The 
solid (dotted) line is the numerical calculation for the inner (outer)
planet averaged over 100 orbits of the outer planet with the planets 
initially aligned with the observer.  
The dashed line is equation (\ref{sigcirc2}).  The large dots are
equations (\ref{reemeqn}) on resonance and equation (\ref{halfres})
halfway between resonances.}
\label{fig3}
\end{figure}

Figure \ref{fig3} shows a numerical calculation of the
standard deviation of the transit timing variations.  We have
used small masses to avoid chaotic behavior since resonant
overlap occurs for $j\ga \mu^{-2/7}$ \citep{wis80}.  Figure \ref{fig4}
zooms in on the 2:1 resonance.  As predicted, the amplitude scales 
as $\epsilon^{-1}$ (equation \ref{wings}), and then steepens to $\epsilon^{-3}$ 
(equations \ref{heavyplanet} and \ref{lightplanet}) closer to 
resonance.  Since the strength of the perturbation is independent
of whether the perturbing planet is interior or exterior,
the strength of the resonances are similar and the shape of the
standard deviation of the transit timing variations
is symmetric about $P_{in}=P_{out}$.
The dashed curve in Figure \ref{fig3} shows the analytic approximation
from equation (\ref{sigcirc2}), which agrees well for $P_{pert}<(2\pi)^{-3/4} P_{trans}$.
The numerical results match the perturbation calculation, equations
(\ref{sigcirc1}) and (\ref{sigcircjason}),
except for near resonance where the change in mean-motion dominates (we have
not bothered to overplot the perturbation calculation since it is
indistinguishable from the numerical results).

There is a dip in $\sigma_2$ near $P_{out}=2.5 P_{in}$ which 
occurs because the amplitude of the timing differences due to
the orbit of the star about the barycentre (eqn. \ref{sigcirc2}) are 
opposite in sign and
comparable in amplitude to the differences due to the perturbation of
the outer planet by the inner planet (eqn. \ref{sigcircjason}).

The analysis in this section breaks down near each resonance because we have
not considered changes to the orbital elements of the perturbing planet.  
In the next section we consider what happens to planets trapped in
resonance.
  
\section{Perturbations for two planets in mean-motion resonance}

The analysis in the previous section assumes that the perturbation 
to the orbit of each planet is
small, so that the interaction can be calculated using the unperturbed
orbits (linear perturbation theory).  This is clearly not the
case near a mean-motion resonance.  We investigate the case of low, initially
zero, eccentricity where we found the standard analyses of this case
\citep[e.g.][]{mur99} to be incorrect.  Here we provide a physically motivated, order
of magnitude, derivation of the perturbations and the transit timing
variations for two planets in a first-order mean-motion resonance.
A rigorous derivation is left for elsewhere, but 
we verify our findings with numerical simulations.

Consider a first order, $j$:$j$+$1$, resonance where the lighter
planet is a test particle.  Qualitatively, the physics of low
eccentricity resonance is as follows: on the nominal resonance, the
two planets have successive conjunctions at exactly the same longitude
in inertial space. The strong interactions that occur at conjunctions 
build up the eccentricity of the test particle and cause a change in semimajor
axis and period. The change in period of the test particle 
causes the longitude of conjunction to drift. Once the
longitude of conjunction shifts by about $\pi$ relative to the
original direction, the eccentricity begins to decrease making a
libration cycle. The libration of the semi-major axes causes
the timing of the transits to change.

This qualitative discussion leads directly to an estimate of the
drifts in transit times. Within each libration cycle the longitude of
conjunction shifts by about half an orbit, mostly due to the period
change of the lighter planet. Since conjunctions occur only once every
$j$ orbits the largest transit time deviation of the lighter planet during
the period of libration is $P/j$
(in this order of magnitude derivation we ignore factors of order 
unity, and take the limit 
of large $j$ so that $j\simeq j+1$ and $P_2\simeq P_1$). 
The
observationally more interesting case is probably that in which the
heavier planet is the transiting one. 
Then, conservation of energy
for the orbiting planets implies that the change in periods is inversely
proportional to the masses, therefore the timing variations are given by
$(m_{light}/m_{heavy})P/j$.  We find an excellent fit to the data for
\begin{equation} \label{reemeqn}
\delta t_{max} \sim {P\over 4.5j}{m_{pert}\over m_{pert}+m_{trans}}.
\end{equation}
The calculations shown in Figure \ref{fig3} verify this analytic scaling 
with $j$.
  
Calculating the libration period is a little more
complicated, but still straightforward. Suppose the period of the test
particle deviates from the nominal resonance by a small fraction
$\epsilon$. Then, consecutive conjunctions drift in longitude by about 
$2\pi j^2 \epsilon$. The number of conjunctions, $N_c$,
before a drift of order $\pi$ in the longitude of conjunctions
accumulates is $N_c \sim j^{-2}\epsilon^{-1}$.  We now estimate $\epsilon$
indirectly.  The test particle gains an eccentricity of order $j^2\mu$ in
each conjunction due to the radial force from the massive planet (this
can be computed from the impulse approximation and the perturbation
equation for eccentricity).  The eccentricity given 
in $N_c$ conjunctions is then of order $\Delta e \sim \mu \epsilon^{-1}$.  
Using the Tisserand relation, the fractional change in semimajor axis associated 
with this change in eccentricity is $j \mu^2 \epsilon^{-2}$.  
Since this is also the fractional change in the period
we have $\epsilon \sim j^{1/3} \mu^{2/3}$ and a libration period of
\begin{equation}\label{plib}
P_{lib} \sim 0.5 j^{-1} \epsilon^{-1} P \sim 0.5 j^{-4/3} \mu^{-2/3}P.
\end{equation}

\begin{figure}
\centerline{\psfig{file=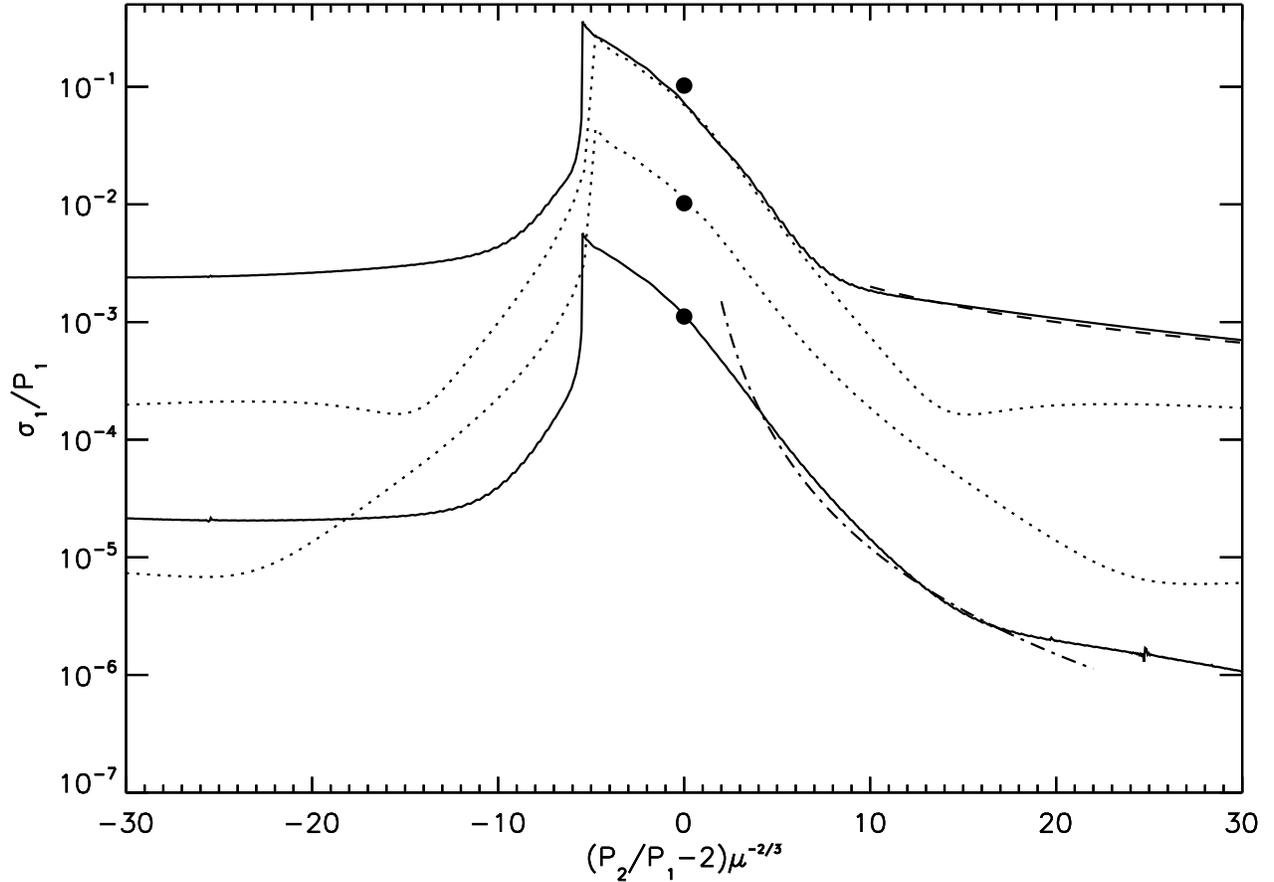,width=\hsize}} 
\caption{Amplitude near the 2:1 resonance versus the difference in 
period from exact resonance for two systems:  one with
$m_1=10^{-5}m_0$ (top solid) and $m_2=10^{-3}$
(lower solid), and the other with $m_1=10^{-6}m_0$ (top dotted) and $m_2=10^{-5}$ 
(lower dotted).  The large dots are equation (\ref{reemeqn}).
The dashed line is equation (\ref{wings}), while the dash-dot line is 
equation (\ref{heavyplanet}).}
\label{fig4}
\end{figure}

We numerically computed the amplitude and period of the transit timing 
variations at the 2:1 resonance.  Figure \ref{fig4} shows a plot of the 
amplitude of the timing variations versus the mass ratio of the
perturbing planet to the transiting planet.  As predicted, the amplitude is of
order the period of the transiting planet when the transiting planet is lighter, 
and varies as the mass ratio when the transiting planet is heavier. 
The libration period measured from the numerical simulations shows the 
predicted behavior, scaling precisely
as $\mu^{-2/3}$ for the more massive planet (with a coefficient of 
$\sim 0.7$ for $j=1$ and $0.5$  for $j>1$ in equation \ref{plib}).
We have compared the numerical values of the amplitude
and period of libration on resonance as a function of $j$.  Despite the
fact that the above scalings were derived in the large-$j$ limit, the
agreement is better than 10 per cent for $j\ge 2$, and accurate to about 40
per cent for $j=1$.

Figure \ref{fig4} shows the more detailed
behavior of the amplitude near the 2:1 resonance.  The amplitude is
maximum slightly below resonance at the location of the cusp.  This may
be understood as follows:  since the simulations are started with
$e_1=e_2=0$, after conjunction the eccentricity grows and the outer planet
moves outwards, while the inner planet moves inward.  This causes
the planets to move closer to resonance, causing a longer time between
conjunctions, leading to a larger change in eccentricity and semi-major
axis.  The cusp is the location where the planets reach exact resonance
at the turning point of libration, at which point $\delta t$ is maximum.
To the right of the cusp, the libration causes the planets to overshoot
the resonance, so the change in eccentricity and semi-major axis is 
somewhat smaller, and hence the amplitude is smaller.
Figure \ref{fig4} shows that the width of the resonance scales as
$\mu^{2/3}$ (the horizontal axis has been scaled with $\mu^{-2/3}$ so
that the curves overlap), so for larger mass planets
the resonant variations have a wider range of influence than the non-resonant
variations discussed in the previous section.  The curves in Figure \ref{fig4}
demonstrate that on-resonance the amplitude scales as $min(1,\mu_{pert}/
\mu_{trans})/j$, while off-resonance the amplitude scales as $\mu_{pert}$.

\section{Non-zero eccentricities}

\begin{figure}
\centerline{\psfig{file=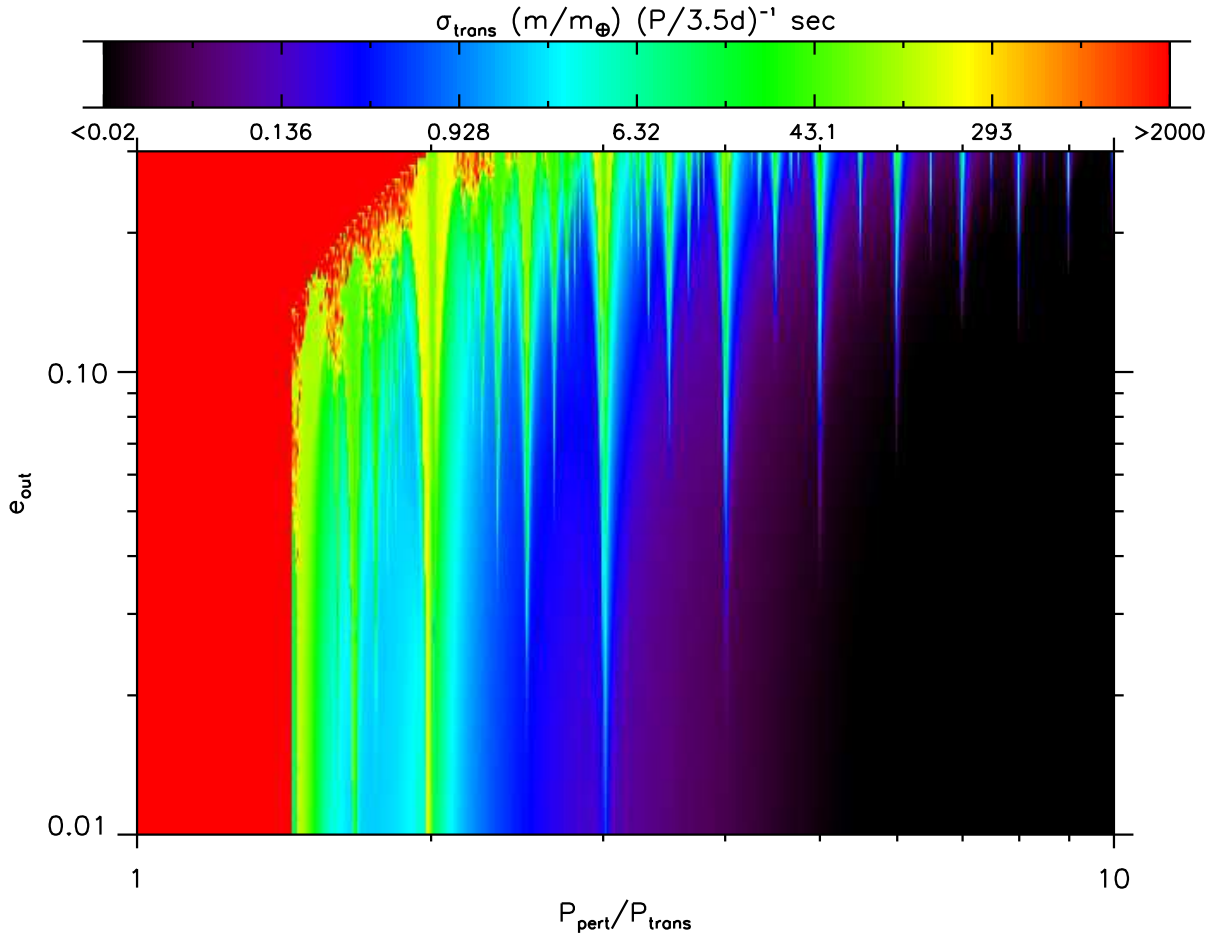,width=\hsize}} 
\caption{Dispersion, $\sigma_{trans}$, of timing variations of HD 209458b due to 
perturbations induced by an exterior earth-mass planet with various 
eccentricities and periods.  The color bar has $\sigma_{trans}$ for a
planet of mass $m_\oplus$ and for a transiting period
of 3.5 days.  Near the period ratio of 4:3 the 
system becomes chaotic.  The increase in the number of resonance peaks, 
particularly those beyond the 2:1 resonance, are from higher order 
terms in the expansion of the Hamiltonian which were truncated for
the near-circular case.}
\label{fig5}
\end{figure}

When either eccentricity is large enough, higher order resonances
become important.  In particular, the resonances that are 1:$m$ begin
to dominate as the ratio of the semi-major axes becomes large; as
the eccentricity of the outer planet approaches unity these resonances
become as strong as first order resonances \citep{pan04}.  Figure
\ref{fig5} shows the results of a numerical calculation where the
transiting planet HD 209458b, with a mass of approximately 0.67 Jupiter
masses, is perturbed by a $1{\rm M_\oplus}$ planet with various eccentricities
(we have taken HD 209458b to have a circular orbit).
Near the mean-motion resonances the signal is large enough that an 
earth-mass planet would be detectable with current technology. The 
amplitude increases everywhere with eccentricity.  This graph can 
be applied to systems with other masses and periods as the timing 
variation scales as $\delta t \propto P_{trans} m_{pert}$ (except
for planets trapped in resonance).

When both planets have non-zero eccentricity, the parameter
space becomes quite large:  the 4 phase space coordinates for each planet,
assuming both are edge-on and 2 mass-ratios give 10 free parameters.
On resonance, the analysis remains similar to the circular case.
The libration amplitude will still be of order $\sim Pj^{-1}$ for the 
lighter planet and $\sim Pj^{-1}(\mu_{light}/\mu_{heavy})$ 
for the heavier planet. However, 
the period of libration will decrease significantly as the eccentricity
increases since $P_{lib} \simeq e^{-1/2} \mu^{-1/2}$.

On the secular time-scale, the precession of the orbits will lead to a
significant variation in the transit timing \citep{mir02}.  
The period of precession, $P_\nu$, may be driven
by other planets, by general relativistic effects, or by a non-spherical
stellar potential, but leads to a magnitude of transit timing
deviation which just depends on the eccentricity for $P_\nu \gg P$.
\citet{mir02} showed that the maximum deviation for $e \ll 1$
is given by
\begin{equation}\label{maxprecession}
\delta t = \frac{eP}{\pi}
\end{equation}
and the timing variations vary with a period that is equal to the period of precession.
For arbitrary eccentricity, the maximum deviation is
\begin{equation}
\delta t = {P \over 2 \pi}\left[\sin^{-1}{y}
+\sin^{-1}{z} +\sqrt{2x-x^2-x^4}\right],
\end{equation}
where $x=(1-e^2)^{1/4}$, $y=(1-x)/e$, and $z=(1-x^3)/e$ (this is derived from the Keplerian solution with a
slowly varying $\varpi$).  This 
approaches $P/2$ as $e\rightarrow 1$.
Typically the eccentricity will vary on the secular time-scale,
so these expressions only hold as long as the variation in $e$ is
much smaller than its mean value.

Rather than systematically studying the entire parameter
space, we now investigate several specific cases of known extrasolar 
planets to demonstrate that detection of this effect should be possible 
once a transiting multi-planet system is found.  Most of these systems
have non-zero eccentricities and several are in resonance, causing
a significant signal.  We summarize the amplitude of the signals
of most known multi-planet systems, if they were seen edge-on, in 
Table 1 (in some cases 
other planets are present which would cause additional perturbations).

\begin{table}
\caption{Timing variations for known multi-planet systems}\label{tab1}
\begin{tabular}{@{}lcccc}
System & $P_{in}$ (d) & $P_{out}/P_{in}$& $\sigma_1$ & $\sigma_2$\cr
\hline
55 Cnc e, b      &  2.81  &  5.21  &  10.5 s  &  2.68 s  \cr
55 Cnc b, c      &  14.7  &  3.02  &  1.61 h  &  14.7 h  \cr
Ups And b, c     &  4.62  &  52.3  &  1.30 s  &  1.61 min  \cr
Gliese 876       &  30.1  &  2.027 &  1.87 d  &  14.6 h  \cr
HD 74156         &  51.6  &  39.2  &  4.98 min  &  42.4 min  \cr
HD 168443        &  58.1  &  29.9  &  12.9 min  &  2.62 h  \cr
HD 37124         &  152   &  9.81  &  3.43 d  &  11.2 d  \cr
HD 82943         &  222   &  2.00  &  34.9 d  &  30.7 d  \cr
PSR 1257+12 b, c &  66.5  &  1.48  &  15.2 min  &  22.3 min  \cr
Earth/Jupiter    &  365   &  11.9  &  1.42 min  &  24.1 s  \cr
\hline
\end{tabular}
\end{table}

\begin{figure}
\centerline{\psfig{file=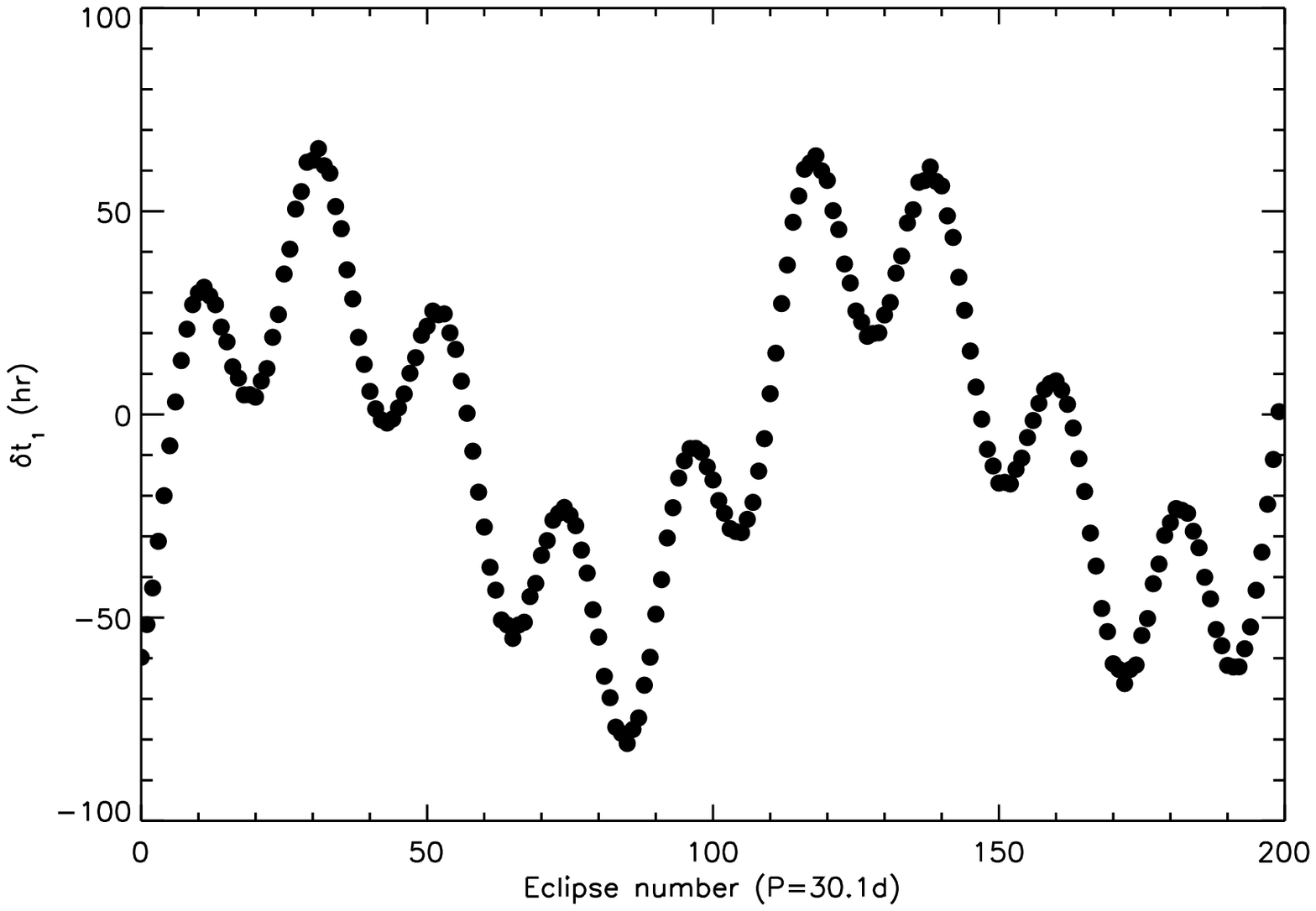,width=5in}} 
\centerline{\psfig{file=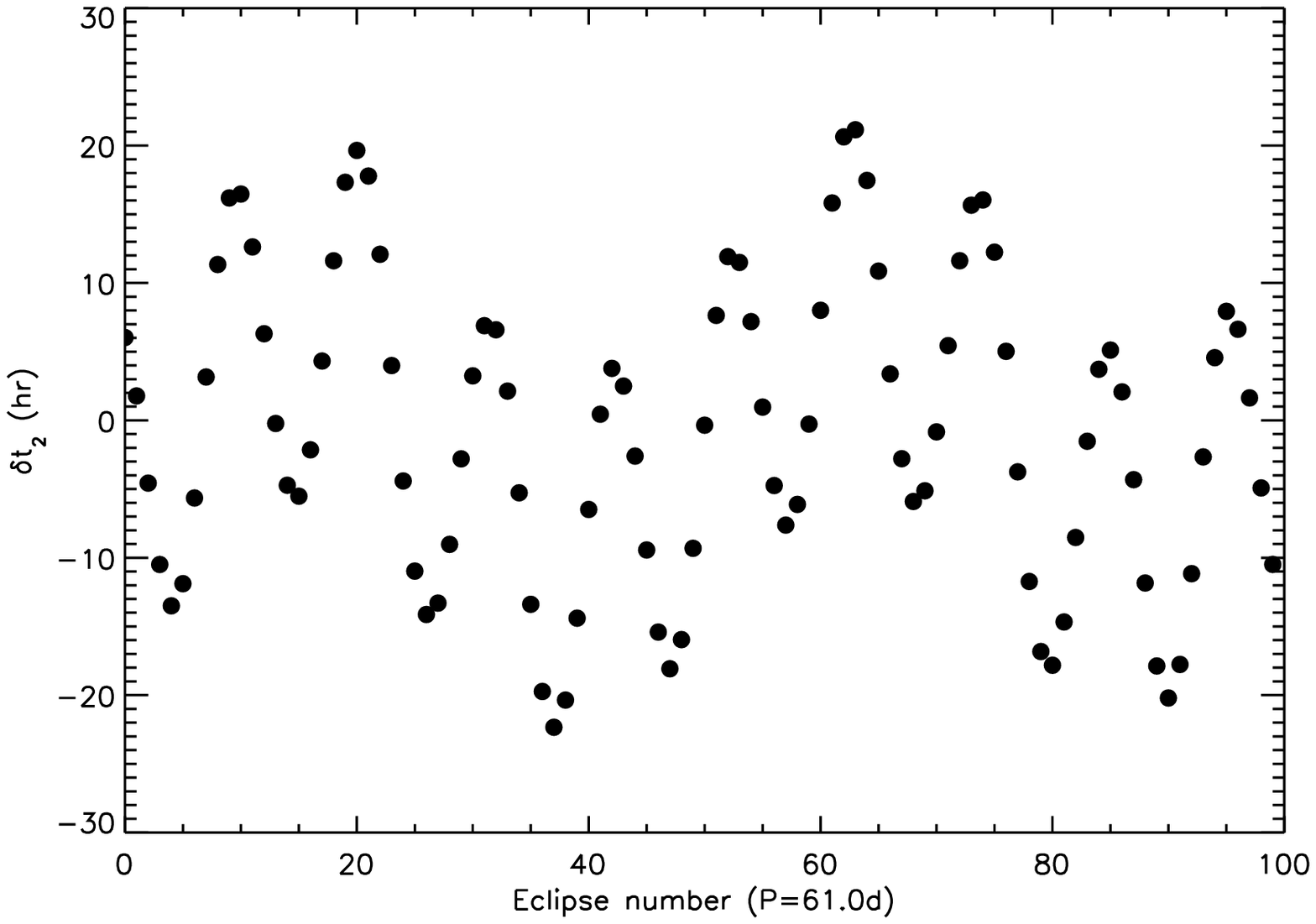,width=5in}} 
\caption{(a) Transit timing variations for Gliese 876 B and
(b) Gliese 876 C.  The vertical axis is in units of hours, while
the horizontal axis is in units of the period of the transiting planet
(given in parentheses for reference).}
\label{fig6}
\end{figure}

The extrasolar planetary system Gliese 876 contains two Jupiter-mass
planets on modestly eccentric orbits which are near the 2:1 mean-motion resonance, 
$P_1=30.1$d and $P_2=61.0$d \citep{mar01}.  Due to the small size of the
M4 host star, the inner planet has a 1.5 per cent probability of transiting
for an observer at arbitrary inclination.  The orbital motion involves
both mean-motion resonance as well as a secular resonance in which the
planets librate about their apsidal alignment.  The apsidal alignment is
in turn precessing at a rate of $-41^{\circ}$ per year \citep{lau04,nau02,
riv01, lau01}.  Figure \ref{fig6} shows the predicted timing variations
if this system were seen edge-on and if the planets are coplanar using
the orbital elements from \citet{lau04}.

The two most prominent periodicities in Figure \ref{fig6} are those associated 
with the 
2:1 libration, with a period of roughly 600 days \citep[20 orbits of the inner 
planet,][]{lau01},
and the long term precession of the apsidal angle with a period of about 3200
days (110 orbits of the inner planet, corresponding to $-41^{\circ}$ yr$^{-1}$).
Evaluating equation (\ref{maxprecession}) gives a peak
amplitude of 1.4 days for the inner planet and 18 hours for the outer
planet which both compare well with the numerical 
results given that the eccentricities are not constant.

The extrasolar planetary system 55 Cancri contains a set of planets, $b$ and $c$,
near the $3$:$1$ resonance having 15 and 45 day periods.
There is some evidence for another planet, $d$, in an extremely long orbit,
and recently a fourth low mass planet, $e$, was found with a 2.8 day period
\citep{mca04}.  The planets $e$, $b$, and $c$ have transit probabilities of 
12, 4, and 2 per cent, respectively, for an observer at arbitrary inclination.  
The orbit of planet $b$ is approximately circular 
while planet $c$ is somewhat eccentric \citep{mar02}.  Table
1 gives the amplitude of the variations for the planets.
We have ignored planet $e$; however, it is at
a large enough semi-major axis to produce a $\sim$ 22 second variation 
due to light-travel time as the barycenter of the inner binary orbits
the barycenter of the triple system were the inner planets transiting.

The double planet system Upsilon Andromedae has a semi-major axis ratio 
of 14 which is not in a mean-motion resonance \citep{but99,mar01}.  The 
inner planet has a short period of 4.6 days, and thus a significant
probability of transiting of about 12 per cent, but has variations which are 
too small to currently be detected from the ground or space.  The outer 
planet has much larger transit timing variations due to its smaller 
velocity, but a much smaller probability of transiting.

The planetary system HD 37124 has two planets with a period ratio of
$\sim 10$ and a period of the inner planet of 241 days \citep{vog00}.  
The outer planet is highly eccentric, $e_2 =0.69$, and so its periapse 
passage produces a large and rapid change in the transit timing of the 
inner planet.
If this system were transiting, the variations would 
be large enough to be detected from the ground.  HD 82943 is in a 2:1
resonance giving variations of order the periods of the planets.
The pulsar planets are near a 3:2 resonance, which would cause large
transit timing variations were they seen to transit the pulsar progenitor
star.  Finally, alien civilizations observing transits of the Sun by 
Jupiter would have to have $\sim$ 10 second accuracy to detect the effect
of the Earth.

\section{Applications}

\subsection{Detection of terrestrial planets}\label{tpdet}

The possibility of detecting terrestrial planets using the transit
timing technique clearly depends strongly on (1) the period
of the transiting planet; (2) the nearness to resonance of
the two planets; (3) the eccentricities of the planets.  
The detectability of such planets also depends on
the measurement error, the intrinsic noise due to stellar variability,
and the number of transit timing measurements.  One requirement for 
the case of an external perturbing planet is that observations should be 
made over a time longer than the period of the timing variations, which
can be longer than the period of the perturbing planet.
Ignoring these complications, a rough estimate of detectability can be 
obtained from comparing the standard deviation of the transit timing with 
the measurement error.

It is worthwhile to provide a numerical example for the case of a hot Jupiter 
with a 3 day period that is perturbed by a lighter, exterior planet on a 
circular orbit in exact 2:1 resonance.  The timing deviation amplitude is of 
order the period (3 days) times the mass ratio (300) or about 3 minutes
(equation \ref{reemeqn}):
\begin{equation}
\delta t=3\left({m_{pert} \over m_\oplus}\right) {\rm min}.
\end{equation}
These variations accumulate over a time-scale of order the period (3 days) times
the planet to star mass ratio to the power of $2/3$, which for a
transiting planet of order a Jupiter mass is about 5 months (equation \ref{plib}):
\begin{equation}
t_{cycle}= 150\left(m_{trans} \over m_J \right)^{-2/3} {\rm days}.
\end{equation}
Such a large signal should easily be detectable from the ground.
With relative photometric precision of $10^{-5}$ from space or from
future ground-based experiments, less massive objects or 
objects further away from resonance could be detected.  The observations could be
scheduled in advance and require a modest amount of observing time with the
possible payoff of being able to detect a terrestrial-sized planet.

\subsection{Comparison to other terrestrial planet search techniques}

To attempt a comparison with other transit timing techniques, we have
estimated the mass of a planet which may be detected at an amplitude
of 10 times the noise for a given technique.  We compare three
techniques for measuring the mass of planets: (1) radial velocity
variations of the star; (2) astrometric measurements; (3) transit
timing variations (TTV).  We assume that radial velocity measurements have
a limit of 0.5 m/s, which is about the highest accuracy that has been
achieved from the ground, and may be at the limit imposed by stellar
variability \citep{but04}.  We assume that astrometric measurements
have an accuracy of 1 $\mu$arcsecond which is the accuracy which
is projected to be achieved by the {\it Space Interferometry Mission}
\citep{for03,soz03}.  Finally, we assume that the transit timing
can be measured to an accuracy of 10 seconds, which is 
the highest accuracy of transit timing measurements of HD 209458 
\citep{bro01}. 

\begin{figure}
\centerline{\psfig{file=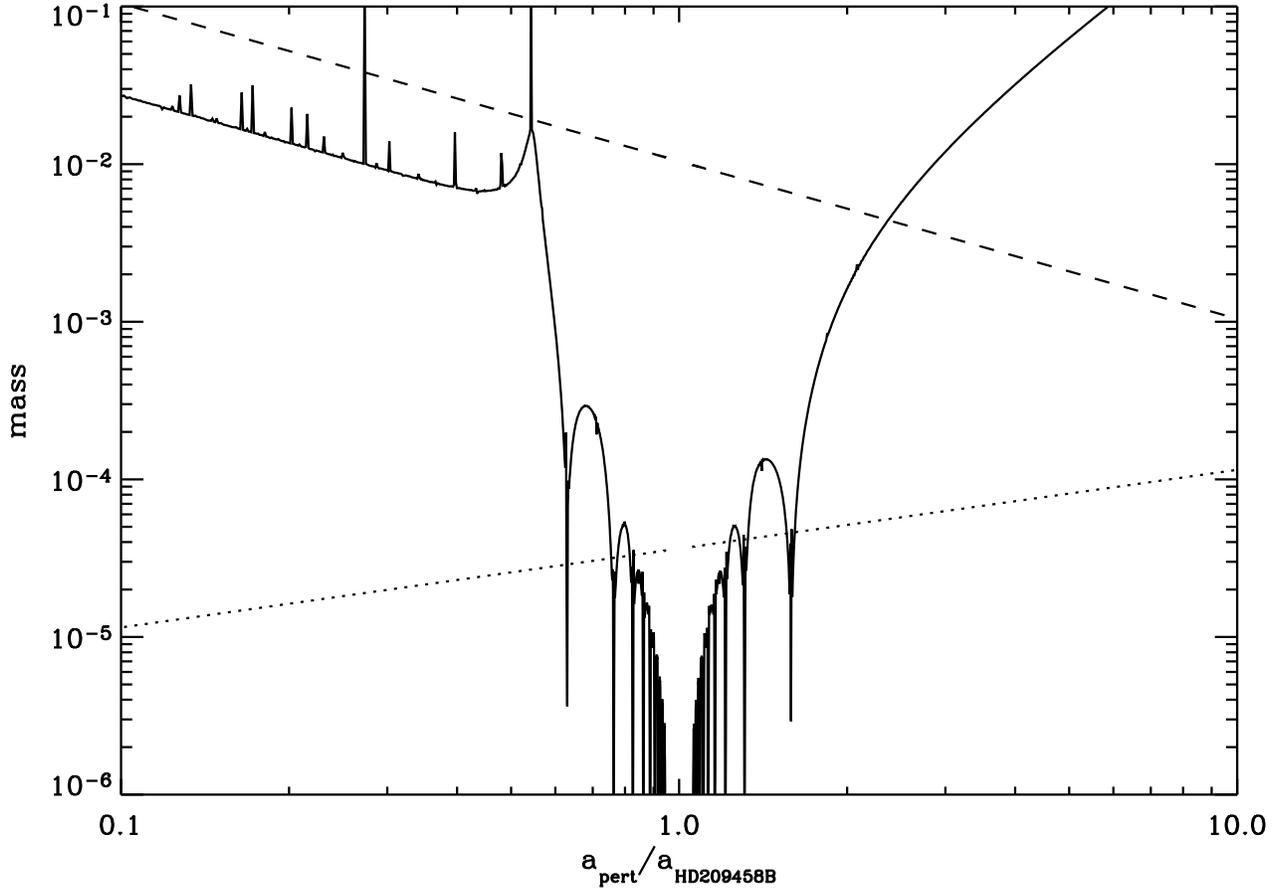,width=\hsize}} 
\caption{Mass sensitivity of various planet detection techniques to
secondary planets in HD 209458.   The vertical axis is the perturbing
planet's mass in units of ${\rm M_\odot}$.   The horizontal axis is
the period ratio of the planets.  The solid line is for the transit timing
technique, the dashed line is astrometric, and the dotted line is the radial
velocity technique.  The large dots are equation (\ref{reemeqn}).}
\label{fig7}
\end{figure}

We concentrate on HD 209458 since it is the best studied transiting 
planet.  This system is at a distance of 46 pc and has a period of 3.5 days.
Figure \ref{fig7} shows a comparison of the 10-$\sigma$ sensitivity
of these three techniques.  The solid curve is computed for
$m_{trans}=6.7\times 10^{-4} {\rm M_\odot}$ and $m_{pert}=10^{-7} {\rm M_\odot}$
and both planets on circular orbits.  When not trapped in resonance, the 
amplitude of the timing variations scales as $m_{pert}/m_0$, so we scale
the results to the mass of the perturber to compute where the timing
variations are 100 seconds -- this determines the sensitivity.  For 
exact resonance we plot equation (\ref{reemeqn}).
The TTV technique is more sensitive than
the astrometric technique at semi-major axis ratios smaller than
about 2.  Off resonance, radial velocity measurements are the technique
of choice for this system, while on resonance the TTV is sensitive
to much smaller planet masses.  Note that in Figure \ref{fig7} the TTV 
and astrometric techniques have the same slope at small $P_{pert}/P_{trans}$.
This is because the transit timing technique is measuring 
the reflex motion of the host star due to the inner planet, which is
also being measured by astrometry.  The solid curve is an {\it upper limit}
to the minimum mass detectable in HD 209458 since a non-zero eccentricity will
lead to larger timing variations (Figure \ref{fig5}) and thus a smaller 
detectable mass.

\subsection{Breaking the mass-radius degeneracy}\label{mrdeg}

In the case that two planets are discovered to transit their host star,
measurement of the transit timing variations can break the degeneracy
between mass and radius needed to derive the physical parameters of the
planetary system.  This has been discussed by \citet{sea03a} who use 
a theoretical stellar mass-radius relation to break this degeneracy.  We 
provide a simplified treatment here to illustrate the nature of the 
degeneracy and how it can be broken with observations of transit timing
variations.

Consider a planetary system with two transiting planets on circular orbits
which are coplanar, exactly edge-on, and have measured radial velocity 
amplitudes.  We'll assume that the star is uniform in surface brightness
and that $m_1, m_2 \ll m_0$.  We'll also assume that the unperturbed periods 
$P_1, P_2$ can be measured from the duration between transits.
Then there are eight physical parameters of interest which
describe the system: $m_0, m_1, m_2, R_0, R_1, R_2, a_1,$ and $a_2$ where $R_i$ 
are the radii of the star and planets.  Without
measuring the transit timing variations, there are a total of ten parameters
which can be measured:  $K_1, K_2, t_{T1}, t_{T2}, t_{g1}, t_{g2}, \Delta F_1,
\Delta F_2, P_1,$ and $P_2$, where $t_{Tj}$ labels the duration of transit
from mid-ingress to mid-egress, $t_{gj}$ labels the duration
of ingress or egress for planet $j$, $K_j$ are the velocity amplitudes of
the two planets, and $\Delta F_j$ are the relative depths of the transits in
units of the uneclipsed brightness of the star (for planet $j$).  Although 
there are more
constraining parameters than model parameters, there is a degeneracy since
some of the observables are redundant.  All of the system
parameters can be expressed in terms of observables and the ratio of
the mass to radius of the star, $m_0/R_0$,
\begin{eqnarray}
{R_j \over R_0} &=& \Delta F_j^{1/2}\cr
{m_j \over m_0} &=& K_j \left({P_j \over \pi t_{Tj} G}\right)^{1/2}  \left({m_0\over R_0}\right)^{-1/2} \cr
a_j &=& {P_j \over 2 \pi} \left({G \pi t_{Tj} \over P_j }\right)^{1/2} \left({m_0\over R_0}\right)^{1/2} \cr
{R_0 \over a_j} &=& {\pi t_{Tj} \over P_j}\cr
{R_j \over a_j} &=& {\pi t_{gj} \over P_j},
\end{eqnarray}
where $j=1,2$ labels each planet.
From this information alone one can constrain the density of the star 
\citep{sea03a}.  For the simplified case discussed here,
\begin{equation}
\rho_* = {3 P \over \pi^2 G t_T^2}
\end{equation}
for either planet \citep[this differs sligthly from the expression in][since we define
the transit duration from mid in/egress]{sea03a}.
If, in addition, one can measure the amplitude of the transit timing variations
of the outer planet, $\sigma_2$, then this determines the mass ratio.  For the
case that the star's motion dominates the transit timing,
\begin{equation}
{m_1\over m_0} = {2 \sqrt{2} \pi \sigma_2 \over P_2^{1/3} P_1^{2/3}}.
\end{equation}
For other cases, the transit timing amplitude can be computed numerically.
Then, from the above expression for $m_i/m_0$ one can find the
ratio of the mass to the radius of the star
\begin{equation}
{m_0 \over R_0} = {1 \over 8\pi^3 G} {P_1^{7/3} P_2^{2/3} K_1^2 \over t_{T1} \sigma_2^2}.
\end{equation}
Combined with the measurement of the density, this gives the absolute
mass and radius of the star.  This
procedure requires no assumptions about the mass-radius relation for the
host star, and in principle could be used to measure this relation.
If one can also measure transit timing variations for the inner planet,
then an extra constraint can be obtained
\begin{equation}
\sigma_1 = {P_1 m_2 \over m_0} f(\alpha),
\end{equation}
where $f(\alpha)$ is a function derived from averaging equation
\ref{sigcirc1}.  (Note that the phase of the orbits is needed
for this equation, which can be found from the velocity amplitude
curve).  This provides an extra constraint on the system,
and thus will be a check that this procedure is robust.

Clearly we have made some drastically simplifying assumptions which
are not true for any physical transit.  The inclination of
the orbits must be solved for, which can be done from the ratio
of the durations of the ingress and transit and the change in flux,
as demonstrated by \citet{sea03a}.
In addition, limb-darkening must be included, and can
be solved for with high signal-to-noise data as demonstrated by
\citet{bro01}.  Finally, the orbits are not generally
circular, so the parameters $e_j, \varpi_j, \Omega_j, \sigma_j$, which 
can be derived from the velocity amplitude measurements,
should be accounted for.
The general solution is rather complicated and would best be
accomplished numerically, but the degeneracy has a similar
nature to the circular case and can in principle be broken by the transit
timing variations.

\section{Effects we have ignored}

We now discuss several physical effects that we have ignored, which
ought to be kept in mind by observers measuring transit timing
variations.

\subsection{Light travel time} \label{lighttravel}

\citet{dee00} carried out a search for perturbing planets
in the eclipsing binary stellar system, CM Draconis, using the changes
in the times of the eclipse due to the light travel time to measure 
a tentative signal consistent with a Jupiter-mass planet at $\sim 1$ AU 
(their technique would in principle be sensitive to a planet on
an eccentric orbit as well, c.f equation \ref{eccentricouter}).
The ``R\"omer Effect'' due to 
the change in light travel time caused by the reflex motion of the 
inner binary is much smaller in planetary systems than in binary stars
since their masses and semi-major axes are small, having an amplitude
\begin{equation}
t_{star}= {a\over c} {m_p\over m_*} \approx 0.5 {\rm sec} \left({m_p \over M_J}\right) \left({a_2 \over 1 {\rm AU}}\right) 
\end{equation}
where $M_J$ is the mass of Jupiter and $a_2$ is the semi-major axis of
the perturbing planet.  This effect is present in the absence of 
deviations from a Keplerian orbit because the inner binary orbits about
about the center of mass.

There can also be changes in the timing of the transit as
the distance of the transiting planet from the star varies.
In this case, the time of transit is delayed by the light travel time
between the different locations where the planet intercepts the beam of light from
the star.  The amplitude of these variations is smaller than the $\sigma$ 
we have calculated by a factor of $\sim v_{trans}/c$,
where $v_{trans}$ is the velocity of the transiting planet.  So, only
very precise measurements will require taking into account light travel
time effects, which should be borne in mind in future experiments (of course
the light-travel time due to the motion of the observer in our solar system 
must be taken into account with current experiments).

\subsection{Inclination}

We have assumed that the planets are strictly coplanar and exactly edge-on.
The first assumption is based on the fact that the solar system is nearly
coplanar and the theoretical prejudice that planets forming out
of disks should be nearly coplanar.  Small non-coplanar effects will
change our results slightly \citep{mir02},  while large inclination effects 
would require a reworking of the theory.  Since some extrasolar planetary systems 
have been found with rather large eccentricities, it is entirely possible that 
non-coplanar systems will be found as well, a possibility we leave for future 
studies.

The assumption that the systems are edge-on is based on the fact that
a transit can occur only for systems that are nearly edge-on.  For small
inclinations our formulae will only change slightly, but may result in
interesting effects such as a change in the duration of a transit, or even
the disappearance of transits due to the motion of the star about
the barycenter of the system.  On a much longer time-scale (centuries), the 
precession of an eccentric orbit might cause the disappearance of transits
since the projected shape of the orbit on the sky can change.  This possibility
was mentioned by \citet{lau04} for GJ 876.

\subsection{Other sources of timing ``noise''}

Aside from the long term effects that have been ignored there are several 
sources of timing noise that must be included in the analysis of observations 
of transiting systems.  These sources of noise could come from
the planet or the host star.  If the planet has a moon or is a binary
planet then there is some wobble in its position causing a change in 
both the timing and duration of a transit \citep{sar99, bro01}.  A moon or ring 
system may transit before the planet causing a shallower transit to appear earlier 
or later than it would without the moon \citep{bro01,sch03,bar04}.  A large
scale asymmetry of the planet's shape with respect to its center of mass
might cause a slightly earlier or later start to the ingress or end of egress.

Stellar variability could also make a significant contribution to the noise. 
Variations in the brightness of the star might affect the accuracy of the
measurement of the start of ingress and the end of egress, which are the
times that are critical to timing of a transit.
Stellar oscillations can cause variations in the surface
of the Sun of $\sim 100$ km in regions of size $10^{3-4}$ km, which corresponds 
to a one second variation for a planet moving at 100 km/s.

\subsection{Coverage Gaps}

Radial velocities and prior transit lightcurves 
predict the epoch of future transits and to schedule
photometric monitoring for the system of interest. Observational
limitations (e.g. bad weather, equipment failure, scheduling requirements)
will lead to transits being missed, which in turn will cause inaccuracies
in $\sigma$.  Since the signal is periodic, $\delta t (t)$ may be
straightforward to extract with a few missed transits; however, if
the outer planet is highly eccentric, then most of the change in
transit timing may occur for a few transits \citep[e.g. 
HD37124; a similar selection effect occurs in radial velocity searches
as discussed by][]{cum04}. In principle
this effect will average out over long observational intervals; however as
in this context ``long'' may mean several decades or more, it will be
important to evaluate the effect of coverage gaps on detections over a
time-scale of months-years.  We will return to this in detail in future 
work. 

We note in passing that the advent of the new astrometric all-sky surveys
such as Gaia \citep{per01} will provide photometry for $\sim 10^7$
stars $>$ 15 mag, with fewer coverage gaps than ground-based observations;  
we thus expect the detection method by transits alone (section \ref{tpdet}) to
really come into its own over the next two decades. Assuming $\sim$ 0.4
detections (three transits) per 10$^4$ stars \citep[c.f.][]{bro03}, we may
expect transit lightcurves of perhaps $\sim 1000$ exoplanetary systems
over the mission lifetime, greatly aiding the determination of $\sigma$ 
for these systems.  The NASA {\it Kepler} mission will also provide
uniform monitoring of about $10^{2-3}$ transiting gas giants \citep{bor03a},
and if flown, the Microlensing Planet Finder (formerly {\it GEST}) will 
discover $10^{4-5}$ transiting gas giants with uniform coverage \citep{ben03}.

\section{Conclusions}

For an exoplanetary system where one or more planets transit the host star the 
timing and duration of the transits can be used to derive several physical 
characteristics of the system.  This technique breaks the degeneracy between 
the mass and radius of the objects in the system.  The inclination, absolute 
mass, and absolute radii of the star and planets can be found; in principle
this could be used to measure the mass-radius relation for stars that are
not in eclipsing binaries.

In addition, TTV can be used to infer the existence of previously undetected 
planets.  We have found that for variations which occur over several orbital 
periods the strongest signals occur when the perturbing planet is either in a 
mean-motion resonance with the transiting planet or if the transiting planet 
has a long period (which, unfortunately, makes a transit less probable).  
The resonant case is more interesting since the 
probability of a planet transiting decreases significantly as the semi-major 
axis becomes large.  Using the TTV scheme it is possible to detect earth-mass 
planets using current observational technologies for both ground based and 
space based observatories.  Data for several transits of HD 209458 could be 
gathered and studied over a relatively short time due to its
small period.  Once the existence of a second planet is established one can 
predict the times at which it would likely transit the host star.  Follow-up 
observations with {\it HST} or high-precision ground-based telescopes
at those times would increase the likelihood of 
detecting a transit of the second planet.  

If the second planet is terrestrial 
in nature, this transit timing method may be the only way currently to 
determine the mass of such planets in other star systems.  Astrometry is 
another possible technique but it may take a decade of technological development 
before the necessary sensitivity is achieved.  In addition, complementary techniques
are necessary to probe different parts of parameter space and to provide extra
confidence that the detected planets are real, given the likely low signal-to-noise
\citep{gou04}.  For the near future the TTV 
technique may be the most promising method of detecting earth-mass planets
around main sequence stars besides the Sun.

We exhort observers to (1) discover longer period transiting planets \citep{sea03b}
since the signal increases with transiting planet's period; (2) increase the
signal-to-noise of ground based differential photometry \citep{how03} for more
precise measurement of the transit times; and (3) examine their transit
data for the presence of perturbing planets \citep{bro01}.

The treatment of this problem has ignored many effects which we plan
to take into account in a future paper in which we will simulate realistic
lightcurves including noise and to fit the simulated data to derive 
the parameters of the perturbing planet, exploring degeneracies in the 
period ratio.  We will also derive the probability of detecting such systems 
taking into account various assumptions about the formation, evolution, and
stability of extrasolar planets.

\section{Acknowledgements}

We acknowledge discussions with Peter Goldreich, Man-Hoi Lee, Jean Schneider.

\appendix
\section{Perturbation theory treatment of nearly circular orbits}
\label{append1}

In this appendix we consider the case of two planets
whose orbits are nearly circular and whose timing variations
are dominated by changes in eccentricity (see \S \ref{circularpert}).
The timing variations can
be computed from a Hamiltonian as described in \cite{mal93a}.  
We keep terms which are first order in the eccentricity because
mutual perturbations between the planets induces an eccentricity of
order $m_2/m_0$.  To first order in the eccentricities, the
Taylor-expanded Hamiltonian is\footnote{We have corrected equation
(26) in \citet{mal93a} which should have a $-\alpha\cos{\psi}$ in the
second line.}
\begin{eqnarray}
H_{int} =& -{Gm_1m_2 \over 2a_2} \left[2P-2\alpha \cos{\psi}
-e_1c^+_0\cos{\left(\lambda_1-\varpi_1\right)}\right.\cr
+&e_2d^+_0 \cos{\left(\lambda_2-\varpi_2\right)}\cr
-&e_1 \sum_{j=1}^\infty \left(c^-_j+\alpha\delta_{j,1}\right)\cos\left((j+1)\lambda_1-j\lambda_2-\varpi_1\right)\cr
-&e_1 \sum_{j=1}^\infty \left(c^+_{j}-3\alpha\delta_{j,1}\right)\cos\left((j-1)\lambda_1-j\lambda_2+\varpi_1\right)\cr
+&e_2 \sum_{j=1}^\infty d^-_j\cos\left(j\lambda_1-(j-1)\lambda_2-\varpi_2\right)\cr
+&\left.e_2 \sum_{j=1}^\infty \left(d^+_j-4\alpha\delta_{j,1}\right)\cos\left(j\lambda_1-(j+1)\lambda_2+\varpi_2\right)\right],
\end{eqnarray}
where $\delta_{i,j}$ is the Kronecker delta, $c^{\pm}_j=\partial_\alpha b^{(j)}_{1/2}\pm 2jb^{(j)}_{1/2}$, $d^{\pm}_j=c^{\pm}_j+b^{(j)}_{1/2}(\alpha)$,
$\alpha = a_1/a_2$, $b^{(j)}_{1/2}(\alpha)$ is the Laplace coefficient,
\begin{equation}
b^{(j)}_{1/2}(\alpha)= {1 \over \pi} \int_0^{2\pi} d\theta {\cos{j\theta} \over
\sqrt{1-2\alpha\cos{\theta}+\alpha^2}},
\end{equation}
where $\psi=\lambda_1-\lambda_2$,
\begin{equation}
P=P(\psi,\alpha)=\left(1-2\alpha\cos{\psi}+\alpha^2\right)^{-1/2},
\end{equation}
and 
\begin{equation}
\partial_{\alpha} b^{(j)}_{1/2}(\alpha) \equiv \alpha{\partial \over \partial \alpha}b^{(j)}_{1/2}(\alpha).
\end{equation}
This equation includes no secular terms since these are higher order
in the eccentricity.  Note that since we have only included the first order 
terms in the eccentricity, the resonant arguments which appear have ratios 
$j+1$:$j$ and $j$:$j+1$ for the mean longitudes. 

The perturbed semi-major axis is given in \citet{mal93b}, and 
we compute the perturbed eccentricity and longitude of periastron
using $h_1=e_1\sin{\varpi_1}$ and 
$k_1=e_1\cos{\varpi_1}$.  Keeping all the resonance terms that exist to first 
order in the eccentricities gives the equations of motion for $h_1, k_1$,
\begin{eqnarray}
\dot h_1 &=& -{1 \over 2} n_1 \alpha {m_2 \over m_0} 
\left[c^+_0 \cos{\lambda_1} \right.\cr
&+& \left. \sum_{j=1}^{\infty} \left\{\left(c^+_j -3\alpha\delta_{j,1}\right)\cos{[(j-1)\lambda_1-j\lambda_2]} \right.\right.\cr
&+&\left.\left.\left(c^{-}_j +\alpha\delta_{j,1}\right)\cos{[(j+1)\lambda_1-j\lambda_2]}\right\}\right],\cr
\dot k_1 &=& {1 \over 2} n_1 \alpha {m_2 \over m_0} 
\left[c^+_0 \sin{\lambda_1} \right.\cr
&-&
\sum_{j=1}^{\infty} \left\{\left(c^+_j-3\alpha\delta_{j,1}\right)\sin{[(j-1)\lambda_1-j\lambda_2]} \right.\cr
&-&\left.\left.\left(c^{-}_j+\alpha\delta_{j,1}\right) \sin{[(j+1)\lambda_1-j\lambda_2]}\right\}\right].\cr
\end{eqnarray}

To find the change in the transit timing we use the orbital elements
to compute the variation in $\dot\theta_1$.  To first order in $e_1$
\begin{equation}
{\dot\theta_1 \over n_{10}}\approx 1 + {\delta n_1\over n_{10}} + 2 k_1 \cos{\left[n_{10}t+\lambda_{10}\right]}+2h_1\sin{\left[n_{10}t+\lambda_{10}\right]}.
\end{equation}
Since we begin with zero eccentricity, we ignore perturbations to $\lambda$ in 
the $\sin$ and $\cos$ terms in this equation.
As in equation (\ref{eclipsetime})
where $\delta \dot\theta_1 = \dot\theta_1 -n_{10}$, we 
integrate this equation to find
\begin{eqnarray}\label{sigcirc1}
\delta t_1 =& 3{m_2 \over m_0}\alpha {n_1 \over \left(n_1-n_2\right)^2} \left(Q(\psi)-{2\psi K(k) \over \pi(1+\alpha)}-\alpha \sin{\psi}\right)\cr
-& n_1{m_2 \over m_0}\alpha \left[{c^+_0 \over n_1^2}\sin{(\lambda_1-\lambda_{10})}\right.\cr
+&\sum_{j=1}^\infty {c_j^+ -3\alpha\delta_{j,1} \over (j-1)n_1-jn_2} \left({\sin{[j\psi]} \over j(n_1-n_2)}-{\sin{[n_1t+j\psi_0]}\over n_1}\right)\cr
-&\left.\sum_{j=1}^\infty {c_{j}^- + \alpha\delta_{j,1} \over (j+1)n_1-jn_2} \left({\sin{[j\psi]} \over j(n_1-n_2)}-{\sin{[n_1t-j\psi_0]}\over n_1}\right)\right],
\end{eqnarray}
where $n_1$ and $n_2$ are taken at their initial values,
$\lambda_{10}=\lambda_1(t=0)$, $\lambda_{20}=\lambda_2(t=0)$, 
$k=2\sqrt{\alpha}/(1+\alpha)$, $K(k)$ is the complete elliptic integral, 
$Q(\psi)$ is defined in the appendix of \citet{mal93b},
and we have dropped any terms which vary linearly with time.  

\begin{figure}
\centerline{\psfig{file=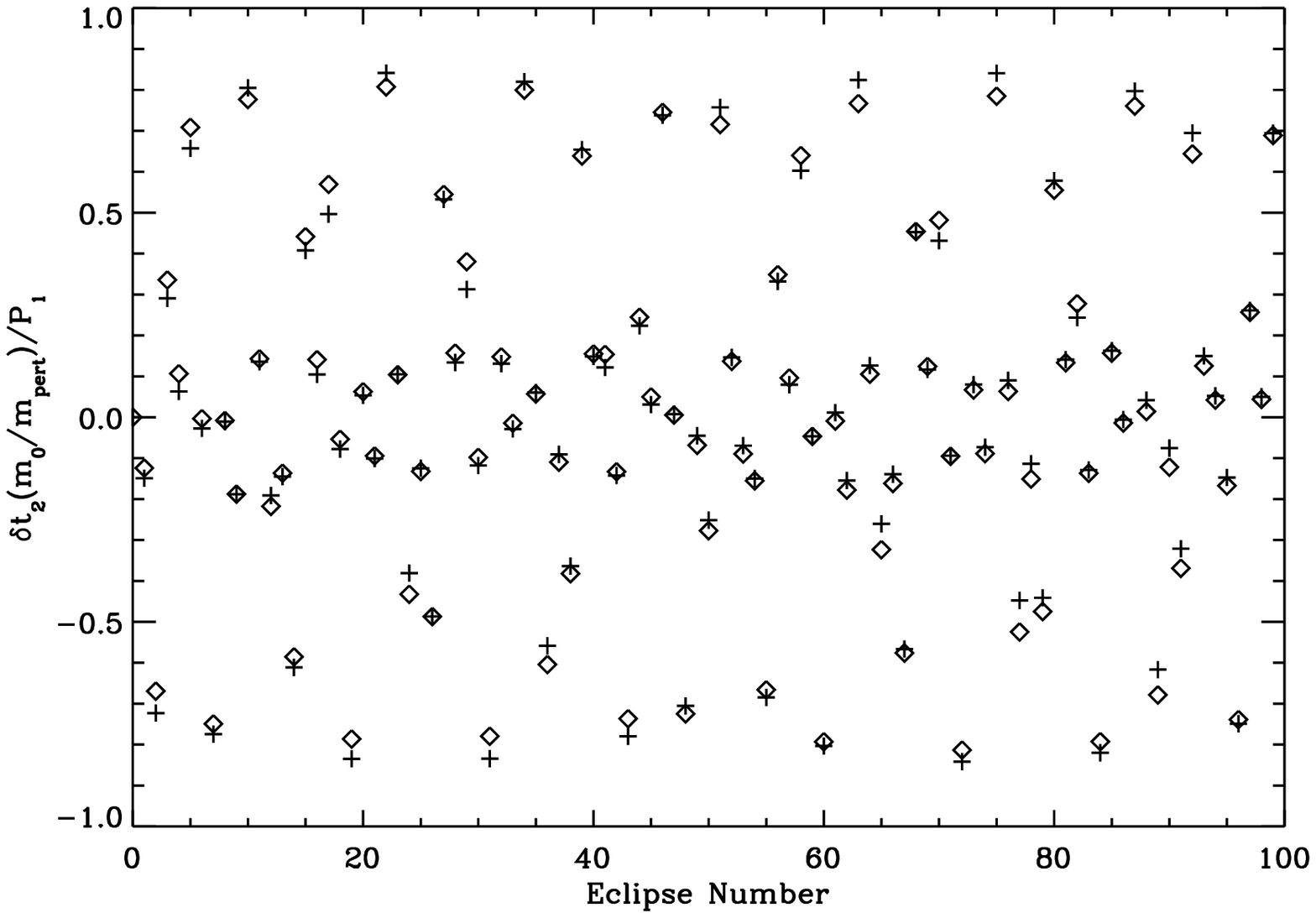,width=\hsize}} 
\caption{Deviations from uniform times between transits for
a transiting outer planet with semi-major axis ratio 1.8. Crosses
represent the analytic result, equation (\ref{sigcircjason}), while
the diamonds are from a numerical integration of the equations
of motion.  The horizontal axis is the number of the transit, while 
the vertical axis shows the timing differences in units of $P_1$,
the period of the inner perturbing planet, multiplied by $m_0/m_{pert}$.}
\label{fig8}
\end{figure}

A similar calculation can be carried
out for perturbations by a planet interior to the transiting planet,
\begin{eqnarray}\label{sigcircjason}
\delta t_2 =& -3 \frac{m_1}{m_0} \frac{n_2}{(n_1-n_2)^2} \left( Q(\psi) - \frac{2\psi K(k)}{\pi(1+\alpha)} - \alpha \sin(\psi) \right) \cr
+& n_2 \frac{m_1}{m_0} \left[\frac{d_0^+}{n_2^2}\sin[\lambda_2-\lambda_{20}] \right. \cr
+& \sum_{j=1}^{\infty} \frac{d_j^+ - 4\alpha\delta_{j,1}}{(jn_1-(j+1)n_2)} \left( \frac{\sin[j\psi]}{j(n_1-n_2)} - \frac{\sin[n_2t+j\psi_0]}{n_2} \right) \cr
-&\left. \sum_{j=1}^{\infty} \frac{d_j^-}{(jn_1-(j-1)n_2)} \left( \frac{\sin[j\psi]}{j(n_1-n_2)} - \frac{\sin[n_2t-j\psi_0]}{n_2} \right) \right]
\end{eqnarray}
A comparison of these equations to numerical calculations is shown in Figure 
\ref{fig8}.  The planets are on initially 
circular orbits with a semi-major axis ratio of 1.8, or period ratio of 2.4.  
The masses are equal and the planets start aligned along the line of sight at the 
first transit.  

In the case that the semi-major axis dominates the timing variations,
one can take perturbed value of the eccentricity ($h$ and $k$) and
compute the change in semi-major axis due to these eccentricities.
The result is a double-series over resonant terms, so we do not
reproduce it here.

\bibliography{agol}
\bibliographystyle{mn2e}
\clearpage

\end{document}